\documentclass{PoS}

\title{Inflation, the scale of supersymmetry breaking, and moduli stabilization}

\ShortTitle{Inflation, the scale of supersymmetry breaking, and moduli stabilization}

\author{\speaker{Emilian Dudas}%
        \\
       Centre de Physique Th\'eorique, Ecole Polytechnique, CNRS, Univ. Paris-Sud, \\91128 Palaiseau Cedex, France\\
       E-mail: \email{emilian.dudas@cpht.polytechnique.fr}}

\author{Clemens Wieck\\
        Departamento de F\'isica Te\'orica UAM and Instituto de F\'isica Te\'orica UAM/CSIC\\
        E-mail: \email{clemens.wieck@uam.es}}

\abstract{We review the effects of heavy scalar fields during inflation in the framework of $\mathcal N = 1$ supergravity. Such heavy scalars can be geometric moduli from string compactifications or stabilizer fields from a different sector of the theory. Even when these fields are heavier than the Hubble scale during inflation, they generically cause backreactions which alter the dynamics of the system. Severe problems may arise when the heavy fields break supersymmetry, which is quite generic for K\"ahler moduli. We illustrate these effects in two examples, chaotic inflation and Starobinsky-like inflation. In chaotic inflation the backreaction of heavy K\"ahler moduli causes a flattening of the quadratic potential. In many setups of Starobinsky-like inflation, however, backreactions spoil the flatness of the plateau.}

\FullConference{18th International Conference From the Planck Scale to the Electroweak Scale \\
		 25-29 May 2015\\
		 Ioannina, Greece }

\begin{document}

\section{Introduction}

Cosmic inflation provides an attractive answer to a series of questions arising in the Hot Big Bang description of the early universe. Its treatment in supersymmetric theories and supergravity has been an active field of research for many years. Moreover, due to the ultraviolet sensitivity of the inflaton potential much work has been devoted to finding a description of inflation in string theory, cf.~\cite{Baumann:2014nda} for a review and a list of references. On the observational side much has been achieved as well, with constraints coming from numerous measurements of the CMB \cite{Ade:2015tva,Ade:2015lrj,Array:2015xqh}. Still, no clear favorite theory of inflation has emerged yet and the construction of fully realistic string theory backgrounds remains challenging. 

However, it is to be expected that heavy scalar fields can matter during an inflationary epoch even if their masses are hierarchically above the Hubble scale, i.e., the dynamical scale of inflation. Such heavy scalars are abundant in string compactifications, and thus a systematic treatment of their backreaction on inflation is desirable. Luckily, many of the backreaction effects of these scalar fields can be treated in four-dimensional effective theories.\footnote{Our approach is  similar in spirit to that of \cite{Rubin:2001in,Dong:2010in}, but we consider more explicit string-effective supergravity theories. For different effects of heavy fields during inflation, cf.~\cite{Achucarro:2010da,Cespedes:2012hu,Achucarro:2012sm}.}
Possible K\"ahler potentials and superpotentials for geometric moduli, especially K\"ahler moduli in type IIB string theory, are known. A number methods are available to stabilize these moduli \cite{Giddings:2001yu,Kachru:2003aw,Balasubramanian:2004uy,Balasubramanian:2005zx,Conlon:2005ki,Westphal:2006tn}. A rather special case, called ``supersymmetric'' or ``strong'' moduli stabilization, is designed to make all F-terms in the moduli sector vanish, and thus to find supersymmetric Minkowski vacua. It has been realized via a racetrack setup in \cite{Kallosh:2004yh} or via the coupling to additional chiral multiplets in \cite{Wieck:2014xxa}. Such so-called stabilizer fields, which are often used in supergravity models to stabilize the inflationary trajectory, are another important class of heavy fields we consider. They, too, can backreact on inflation if supersymmetry is broken spontaneously at a high scale. 

In this work we review recent attempts to combine these stabilization schemes with viable supergravity models of inflation in string-effective Lagrangians. We study the backreaction of heavy K\"ahler moduli on inflation by integrating out the moduli at tree-level. We repeat the same analysis for stabilizer fields in popular models and reveal effects which may not be overlooked in the case of high-scale supersymmetry. The latter results have been published in \cite{Buchmuller:2014pla} for the simplest example of chaotic inflation with a stabilizer field and its deformations. The impact of stabilized K\"ahler moduli on chaotic inflation has been studied in detail in \cite{Buchmuller:2015oma}. Similar setups often arise in string theory embeddings of axion monodromy inflation. A more general study, with applications to Starobinsky-like and natural inflation, has been presented in \cite{Dudas:2015lga}. Before we proceed to discuss the backreaction of the aforementioned heavy scalars, let us recall a few basic facts about inflation in supergravity.

\subsection{Chaotic inflation in supergravity}

One of the simplest models of chaotic inflation \cite{Linde:1983gd} features a quadratic potential for the inflaton field $\varphi$,
\begin{equation}\label{eq:Cha1}
V = \frac12 m^2 \varphi^2\,.
\end{equation}
With a small inflaton mass, $m \approx 6 \times 10^{-6}$ in Planck units, 60 $e$-folds of inflation are observable with $n_{\rm s} \sim 0.97$ and $r \sim 0.13$. The expansion of space is driven by slow-roll of the inflaton field from $\varphi_\star \sim 15$ to its minimum at the origin. Naively we could try to implement this model in supergravity by choosing
\begin{equation}
K=\frac12(\Phi + \overline \Phi)^2\,, \qquad W = \frac12 m \Phi^2\,,
\end{equation}
where $\varphi = \sqrt2 {\rm Im}\, \Phi$ is protected by a shift symmetry. However, a computation of the $\mathcal N = 1$ supergravity scalar potential reveals 
\begin{equation}
V = \frac12m^2 \varphi^2 - \frac{3}{16}m^2\varphi^4\,,
\end{equation}
which is negative for $\varphi \gtrsim 1$ and thus unviable for our purposes. A solution to this problem has been suggested in \cite{Kawasaki:2000yn} in the form of a stabilizer field. Coupling an additional chiral multiplet in the following way,
\begin{equation}\label{eq:Cha2}
K=\frac12(\Phi + \overline \Phi)^2 + |S|^2\,, \qquad W = m S \Phi\,,
\end{equation}
leads to Eq.~(\ref{eq:Cha1}) while $\langle S \rangle = 0$ during inflation.\footnote{Note that a consistent decoupling of $S$ usually requires higher-order terms in $K$. Alternatively, such a decoupling can possibly be achieved by the use of nilpotent superfields, cf.~the discussions in \cite{Antoniadis:2014oya,Ferrara:2014kva,Kallosh:2014via,Dall'Agata:2014oka,Carrasco:2015uma,Carrasco:2015pla,Dudas:2015eha,Hasegawa:2015era}.}

\subsection{Starobinsky-like inflation in supergravity}

One of the first inflation models was proposed by A.~Starobinsky in \cite{Starobinsky:1980te}. The original version features a modification of Einstein gravity,
\begin{equation}
S = \int {\rm d}^4 x \sqrt{-g}\left(\frac12 R + \alpha R^2 \right)\,,
\end{equation}
which is dual to a theory of Einstein gravity coupled to a scalar field $\varphi$,
\begin{equation}\label{eq:Star3}
S = \int {\rm d}^4 x \sqrt{-g}\left[\frac12 R - \frac12 (\partial \varphi)^2 - \frac{1}{16 \alpha}\left(1-e^{-\sqrt{\frac23} \varphi} \right)^2 \right]\,.
\end{equation}
The potential for $\varphi$ is exponentially flat at large field values, which makes it attractive in the eyes of the inflationary model-builder. Since \cite{Starobinsky:1980te} many proposals have been developed to realize similar scenarios in supergravity. One possible Lagrangian reads \cite{Cecotti:1987sa,Ferrara:2013wka}
\begin{equation}\label{eq:Star2}
\mathcal L = \left[ -|S_0|^2 + h\left(\frac{\mathcal R}{S_0},\frac{\overline{\mathcal R}}{\overline S_0} \right) |S_0|^2 \right]_D + \left[ W\left(\frac{\mathcal R}{S_0}\right) S_0^3 \right]_F\,,
\end{equation}
where $S_0$ denotes the chiral compensator superfield, $\mathcal R$ the chiral curvature superfield, and $h$ is a real function. According to \cite{Cecotti:1987sa,Ferrara:2013wka} this can be recast into a two-derivative formulation for two chiral superfields. Writing Eq.~(\ref{eq:Star2}) as 
\begin{equation}
\mathcal L = \left[ -|S_0|^2 + h(C,\overline C) |S_0|^2 \right]_D + \left[ \Lambda \left(C-\frac{\mathcal R}{S_0} \right) + W(C) S_0^3 \right]_F\,,
\end{equation}
where $\Lambda$ ensures $C = \mathcal R / S_0$, we can define $\Lambda = T - \frac12$ and find a different formulation of the same theory,
\begin{equation}
K = -3 \log \left[ T + \overline T - h(C,\overline C) \right]\,, \qquad W = C \left(T -\frac12 \right) + W_0\,.
\end{equation}
In this formulation ${\rm Re}\,T = e^{\sqrt{\frac23} \varphi}$ contains the inflaton field and $C$ plays the role of a stabilizer field. We will come back to this example in Section 2.2.

\section{Constraints from heavy moduli and stabilizer fields}

We are now in a position to discuss the backreaction of heavy moduli and stabilizer fields. Let us begin by considering setups without stabilizer fields which generically arise as low-energy effective theories of many string compactifications. In all of them a number of geometric moduli, additional scalar fields, deserve our attention regarding the evolution of the early universe. In type IIB string theory it is well-known that complex structure moduli and the axio-dilaton field can be stabilized supersymmetrically by fluxes \cite{Giddings:2001yu}. Therefore we assume them to decouple from the remaining dynamics in what follows. K\"ahler moduli, on the other hand, obey a no-scale symmetry which requires breaking by quantum corrections in the effective action in order to render them massive. In the setups of KKLT \cite{Kachru:2003aw}, the Large Volume Scenario \cite{Balasubramanian:2005zx,Conlon:2005ki}, and K\"ahler Uplifting \cite{Balasubramanian:2004uy,Westphal:2006tn}, for example this is achieved by non-perturbative superpotentials or a combination with the leading-order $\alpha'$ correction of $K$. In all of those mechanisms the K\"ahler moduli are stabilized in Minkowski or de Sitter vacua with spontaneously broken supersymmetry. Although their masses can be much larger than the inflationary Hubble scale, a caveat remains: The scale of supersymmetry breaking, parameterized by $m_{3/2}$, must be very large as well. This is to avoid destabilization of the moduli during inflation, since both the mass of the moduli and the barriers protecting their metastable vacua are proportional to $m_{3/2}$. In KKLT we must require $m_{3/2} > H$ \cite{Kallosh:2004yh}, while in LVS even $m_{3/2}/\sqrt{\mathcal V} > H$ \cite{Conlon:2008cj}, where $\mathcal V \gg 1$ denotes the volume of the compact manifold. This is avoided in the strong moduli stabilization schemes mentioned in the Introduction. Since the moduli are stabilized in metastable supersymmetric Minkowski vacua, their barriers are independent of $m_{3/2}$. 

However, even if the above requirements are fulfilled in moduli stabilization schemes which involve spontaneous supersymmetry breaking, the K\"ahler moduli may have inflaton-dependent vacua during inflation and thus backreact on the dynamics. A large class of models is captured by the ansatz
\begin{equation}\label{eq:Gen1}
K = K_0(T_\alpha, \overline T_\alpha) + K_1 (\Phi + \overline \Phi, X, \overline X, T_\alpha, \overline T_\alpha)\,, \qquad W = W_{\rm inf}(\Phi) + W_1(X,T_\alpha)\,.
\end{equation}
Here we assume that the inflaton, contained in $\Phi$, is protected by a shift symmetry. $T_\alpha$ denote a number of moduli fields, while $X$ is responsible for the uplift of AdS vacua which are generically produced by an appropriate choice of $W_1$. Starting from this, our goal is to find the effective inflaton potential after integrating out all heavy modes at tree-level. In the most successful moduli stabilization setups on the market, the biggest impact comes from the lightest K\"ahler modulus. The strategy is thus to treat $W_{\rm inf}$ as a perturbation of the modulus potential, and to expand
\begin{equation}
V = V_0 + V_1 + V_2 + \dots\,.
\end{equation}
Here $V_0$ denotes the modulus potential after inflation, $V_1$ contains terms up to $\mathcal O(W_{\rm inf})$, $V_2$ contains all terms up to $\mathcal O(W_{\rm inf}^2)$, and so on. To compute the various terms we expand the moduli fields around their true minima after inflation has ended,
\begin{equation}
T_\alpha = T_{\alpha,0} + \delta T_\alpha(\Phi)\,.
\end{equation}
Denoting $\rho_\alpha = (T_\alpha, \overline T_\alpha)$ this leads to the following expansions at quadratic order in $\delta T_\alpha$,
\begin{eqnarray}
V_0(\rho_\alpha) &\approx& \Lambda_0^4 + \frac12 \delta\rho_\alpha M_{\alpha \beta}^2 \delta \rho_\beta\,, \\
V_1 (\rho_\alpha, \Phi) &\approx& V_1(\rho_{\alpha,0}, \Phi) + \delta \rho_\alpha \partial_{\rho_\alpha}V_1\,, \\
V_2 (\rho_\alpha, \Phi) &\approx& V_2(\rho_{\alpha,0}, \Phi)\,,
\end{eqnarray}
where $\Lambda_0$ contains possible contributions to a cosmological constant. The expanded potential is minimized by $\delta \rho_\alpha \approx - M_{\alpha \beta}^{-2} \partial_{\rho_\beta} V_1$. Subsequently, we can write the leading-order backreaction term as follows,
\begin{equation}
V_{\rm back} = - \frac12 \partial_{\rho_\alpha} V_1 M_{\alpha \beta}^{-2}  \partial_{\rho_\beta} V_1\,.
\end{equation}
For details of the computation and more explicit expressions we refer the reader to \cite{Dudas:2015lga}. In this review, for the sake of illustration we discuss this backreaction by means of a simple example in Section 2.1.

But first let us consider another class of models, one which is not covered by the ansatz in Eqs.~(\ref{eq:Gen1}). Setups involving stabilizer fields are often captured by
\begin{equation}\label{eq:Gen2}
K = K(\Phi + \overline \Phi, S, \overline S, X, \overline X, T_\alpha, \overline T_\alpha)\,, \qquad W = M S f(\Phi) + W_1(X, T_\alpha)\,,
\end{equation}
where $S$ denotes the stabilizer multiplet, $M$ is a mass scale, and $f$ is a holomorphic function. The virtue of the stabilizer usually lies in the fact that its F-term generates the positive definite inflaton potential, while $S$ itself remains stabilized at the origin. Thus, in the absence of additional ingredients and $W_1$, the stabilizer decouples from the dynamics of inflation, which is driven by the imaginary part of $\Phi$. However, things change once we introduce a piece $W_1$ involving moduli or other fields which break supersymmetry. Whenever $W_1$ does not vanish in the vacuum, there is a mixing of the stabilizer and the inflaton field in the supergravity action. For the above choice of $K$ the imaginary part of $S$ couples to the inflaton as follows,
\begin{equation}
V \supset m_{3/2} g(\varphi) {\rm Im}\, S\,,
\end{equation}
where $m_{3/2}$ is proportional to the vacuum expectation value of $W_1$ and $g$ is a model-dependent function of the inflaton field $\varphi = {\rm Im}\,\Phi $. This linear term displaces $S$ from its minimum at the origin and backreacts on inflation. Inserting the inflaton-dependent minimum back into the action we find 
\begin{equation}
V(\varphi) = |M f(\varphi)|^2 - m_{3/2}^2 \frac{g(\varphi)^2}{m_S^2}\,.
\end{equation}
Here $m_S$ denotes the mass of the stabilizer field. Apart from overall normalization due to the pre-factor $e^K$, the first piece is the original inflaton potential in the absence of supersymmetry breaking, while the second piece corrects the potential. Because it is always negative we can conceive that there is a threshold value of $m_{3/2}$ above which inflation is generically impossible. Naively one may think that the correction may be removed by increasing $m_S$, but this is not possible in most examples. For the example of chaotic inflation, specifically the setup of Eqs.~(\ref{eq:Cha2}), this has been worked out in detail in \cite{Buchmuller:2014pla}. In this review we consider a different example in Section 2.2.

\subsection{Example I: Chaotic inflation}

Let us illustrate the backreaction of a single K\"ahler modulus in the example of chaotic inflation, in a setup without a stabilizer field. To this end, a special case of the Lagrangian defined by Eqs.~(\ref{eq:Gen1}) is
\begin{equation}
K = - 3 \log (T + \overline T) + \frac12 (\Phi + \overline \Phi)^2\,, \qquad W = \frac12 m \Phi^2 + W_{\rm mod}(T)\,.
\end{equation}
This theory has been treated in detail in \cite{Buchmuller:2015oma} for various choices of $W_{\rm mod}$ corresponding to the aforementioned stabilization schemes. In that reference also a possible uplift to a Minkowski background is considered explicitly, something we omit here for the sake of brevity. What is important is that we can use the methods outlined above to compute the leading-order backreaction of $T$ on the potential of the inflaton field $\varphi$. What we find, up to an overall rescaling due to $K$, is the following,
\begin{equation}
V(\varphi) = \frac12 m^2 \varphi^2 + \frac{c}{2} m m_{3/2} \varphi^2 - \frac{3}{16}m^2 \varphi^4 + \dots\,.
\end{equation}
Here $c$ is a constant which depends on the particular choice of $W_{\rm mod}$, $m_{3/2}$ is proportional to the vacuum expectation value of $W_{\rm mod}$ and thus to the mass of $T$, and the dots denoted higher-order corrections suppressed by powers of $T_0$. Apparently we recover the original inflaton potential, which is unfeasible for inflation as discussed in Section 1.1, plus a backreaction term which becomes larger as the modulus becomes heavier. Luckily, as demonstrated in \cite{Buchmuller:2015oma}, $m_{3/2}$ can always be chosen large enough so that this new term drives inflation and dominates over the negative term for a sufficient field range. The resulting effective potential is schematically depicted in Fig.~1.\footnote{As in \cite{Buchmuller:2015oma}, however, we remark that the parameter choices necessary to make the model function are highly questionable from the perspective of the underlying string theory.}
\begin{figure}[t]
\centering
\includegraphics[width=0.75\textwidth]{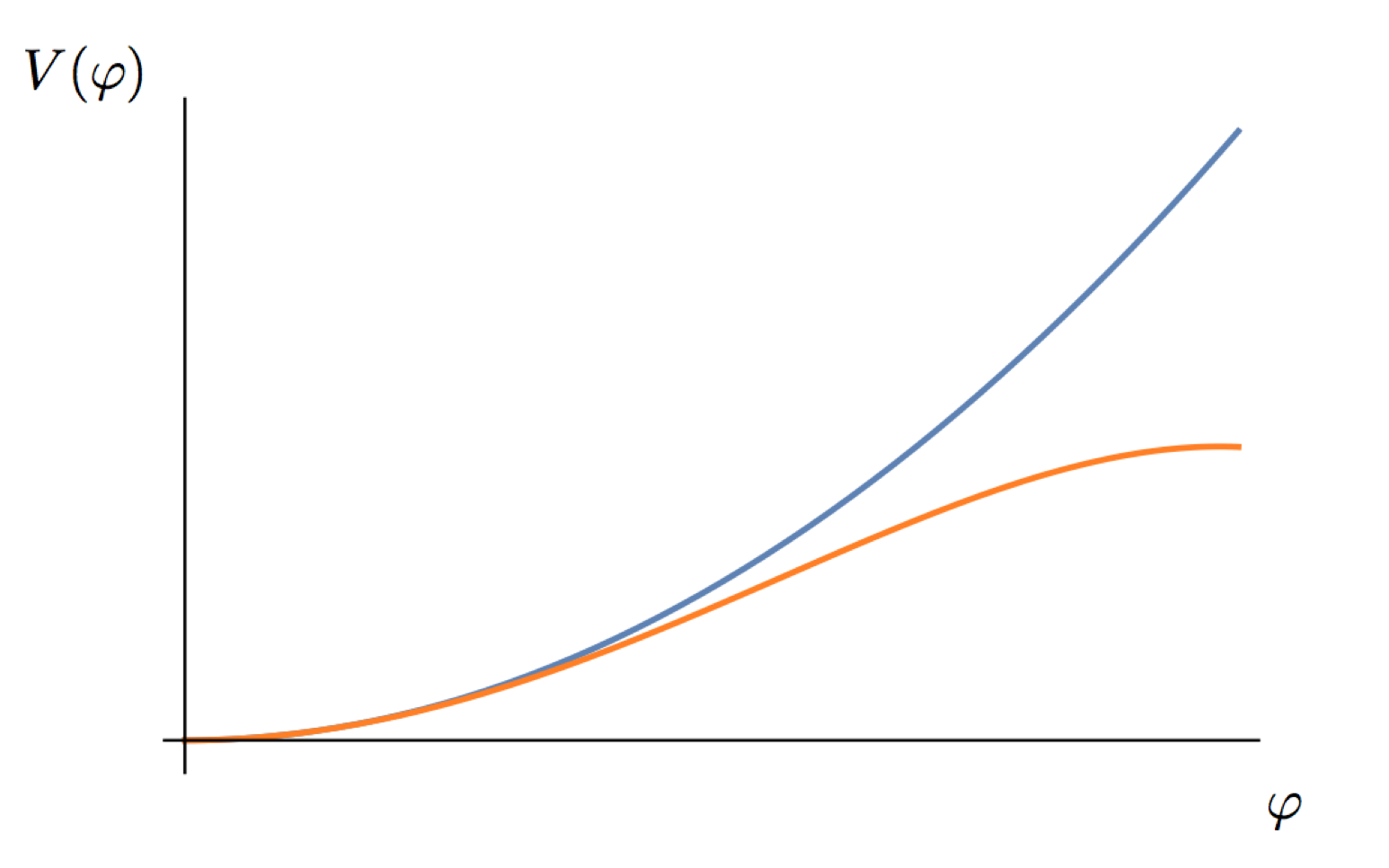} 
\caption{Schematic depiction of the quadratic inflaton potential (blue line) and effective inflaton potential after integrating out $T$ (orange line). The parameters can always be chosen that the model produces 60 $e$-folds of inflation, starting close to the hilltop.}
\end{figure}
We observe the shape of a flattened potential, with a hilltop and subsequent downturn. The latter is far enough out in field space for 60 $e$-folds of inflation to be produced in accordance with observations. 

As illustrated in Fig.~2, the observables $n_{\rm s}$ and $r$ of this model are slightly more appealing than those of pure quadratic inflation. 
\begin{figure}[t]
\centering
\includegraphics[width=0.75\textwidth]{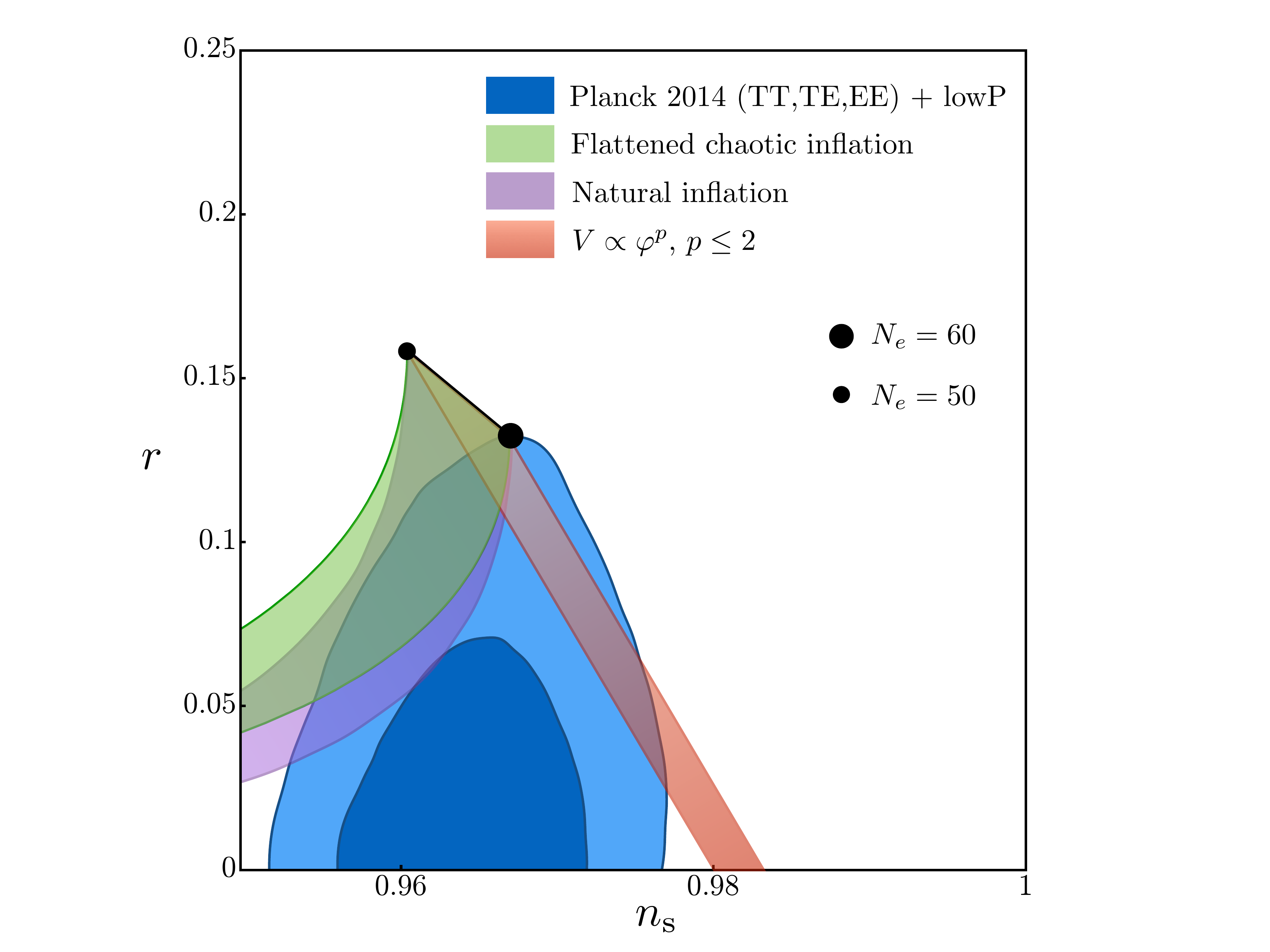} 
\caption{Predicted observables in our setup producing flattened quadratic inflation, compared to those of natural inflation and monomial potentials.}
\end{figure}
As is to be expected, the observables approach those of natural inflation, since the cosine function has exactly the same shape at small field values. The reason why the predictions are not exactly the same is quite subtle. The hilltop of the potential in Fig.~1 can actually never be reached since it corresponds to the point in field space where $T$ is destabilized by the inflationary vacuum energy. This is quite evident in a three-dimensional plot of the two-field system, cf.~Fig.~3. The example chosen is that of KKLT, i.e., $W_{\rm mod} (T) = W_0 + A e^{-a T}$ with a suitable set of parameters, and with an appropriate uplift of the vacuum.
\begin{figure}[t]
\centering
\includegraphics[width=0.85\textwidth]{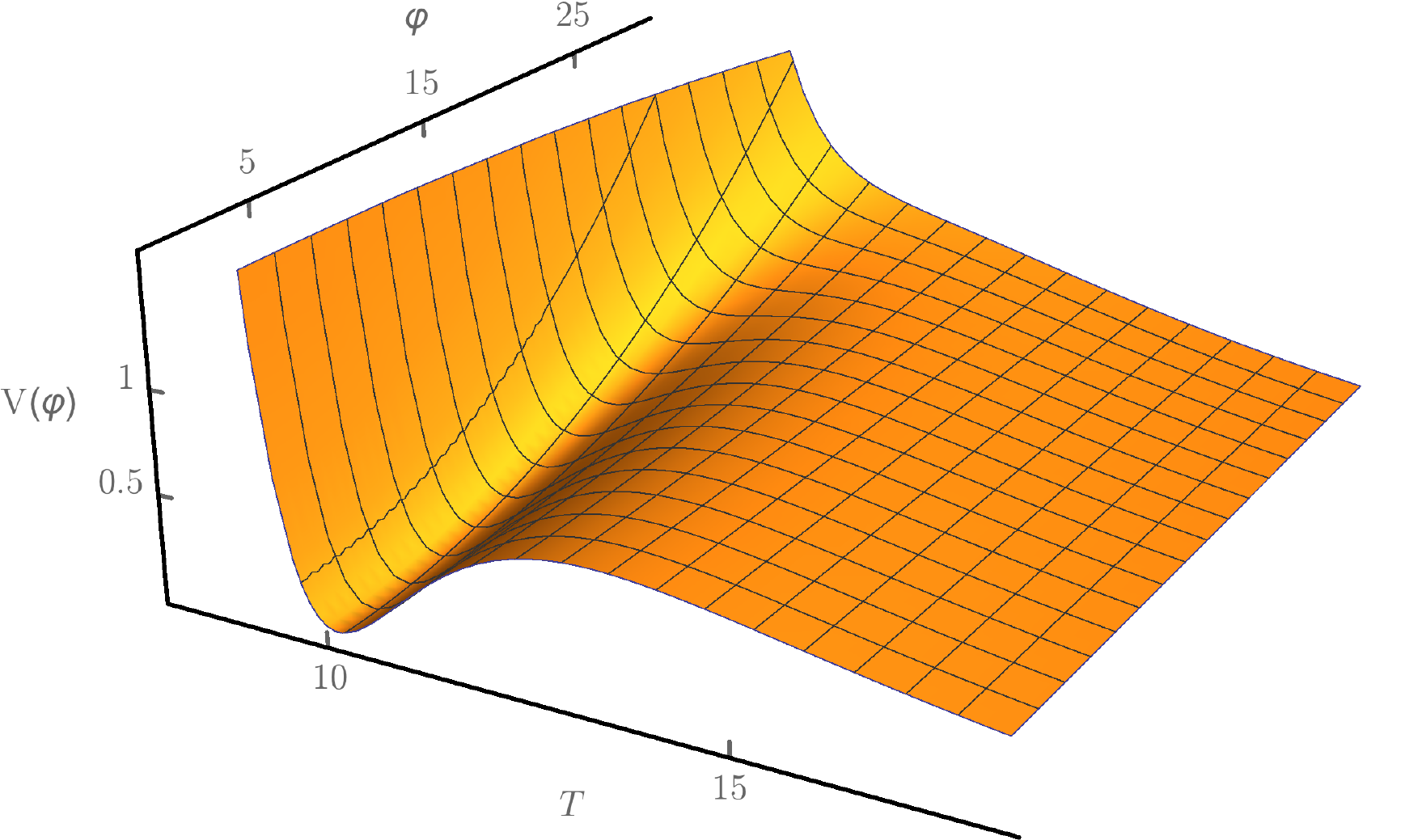} 
\caption{Two-field potential for the example of KKLT. The overall scale of $V(\varphi)$ is arbitrary. Inflation is possible in the valley of the modulus minimum, as long as the initial conditions of the system are such that the latter is not destabilized.}
\end{figure}

Inflation is only possible as long as the initial conditions of the system are chosen such that $T$ remains in its minimum. Then inflation can proceed in the valley depicted in the three-dimensional plot. This choice corresponds to a point slightly to the left of the hilltop in Fig.~1. The condition for this to happen can be translated to a lower bound on the gravitino mass,
\begin{equation}
m_{3/2} > 2 \times 10^{14}\,{\rm GeV}\,,
\end{equation}
in KKLT and slightly higher for other stabilization schemes.

\subsection{Example II: Starobinsky-like inflation}

As we have seen in the previous example, the backreaction of a heavy modulus can be conducive for certain models of inflation. We now turn to an example of the opposite kind: The backreaction of a stabilizer field prohibits inflation in a plateau-like supergravity setup.

Consider the Cecotti model discussed in Section 1.2, in a slightly more convenient notation,
\begin{eqnarray}
K = -3 \log \left( \Phi + \overline \Phi - |S|^2 + \frac{\xi}{3}|S|^4 \right)\,, \qquad W = M S (\Phi-1)\,.
\end{eqnarray}
The quartic term for the stabilizer field was introduced in \cite{Kallosh:2013xya} and is necessary for $S$ to sufficiently decouple during inflation. With $S = 0$ during inflation this produces the Starobinsky potential, cf.~Eq.~(\ref{eq:Star3}). Once we add a piece which breaks supersymmetry in the vacuum, like 
\begin{equation}
W_1 = f X + W_0 \,, 
\end{equation}
where $X$ is a chiral Polonyi field and $f$ and $W_0$ are treated as constants, the picture changes.\footnote{We follow the notation of the general ansatz in Eqs.~(\ref{eq:Gen2}). For simplicity, we do not consider additional moduli fields $T_\alpha$ in this example. Furthermore, we assume that $X$ is stabilized close to the origin and with a large mass by a suitable K\"ahler potential. That could be, for example, a canonical one with a one-loop correction similar to the K\"ahler potential for $S$.} We find a term linear in $S$ which multiplies the gravitino mass, proportional to $W_0$, and which causes a backreaction on the inflaton potential. The leading-order effective potential after integrating out the stabilizer field becomes\footnote{Once more we refer to \cite{Dudas:2015lga} for details.}
\begin{equation}
V(\varphi) = \frac{M^2}{12} \left(1-e^{-\sqrt{\frac23}\varphi}\right)^2 + \frac{f^2}{8} e^{-3\sqrt{\frac23}\varphi} - \frac{9 W_0^2}{8 \xi} e^{-4\sqrt{\frac23}\varphi}\,.
\end{equation}
As expected, in the supersymmetric limit $f=W_0=0$ we recover the uncorrected Starobinsky potential. If the scale of supersymmetry breaking is small, $f \ll M$, inflation proceeds as desired and ends in a non-supersymmetric Minkowski vacuum if $ f \sqrt{\xi}\approx \sqrt3 W_0$. As $f$ and $W_0$ are increased, the latter relation must be modified for inflation to end in a vacuum with parametrically small cosmological constant. This works until a certain threshold. Specifically, for 
\begin{equation}
W_0 > \frac19 \sqrt{\frac12 + \frac{1}{\sqrt3}} \ M\,,
\end{equation}
the effective potential does not admit Minkowski vacua, or de Sitter vacua with a small cosmological constant, at all.\footnote{Here we have assumed that $\xi \sim \mathcal O(1)$, which is not particularly restrictive.} This situation is schematically depicted in Fig.~4.
\begin{figure}[t]
\centering
\includegraphics[width=0.75\textwidth]{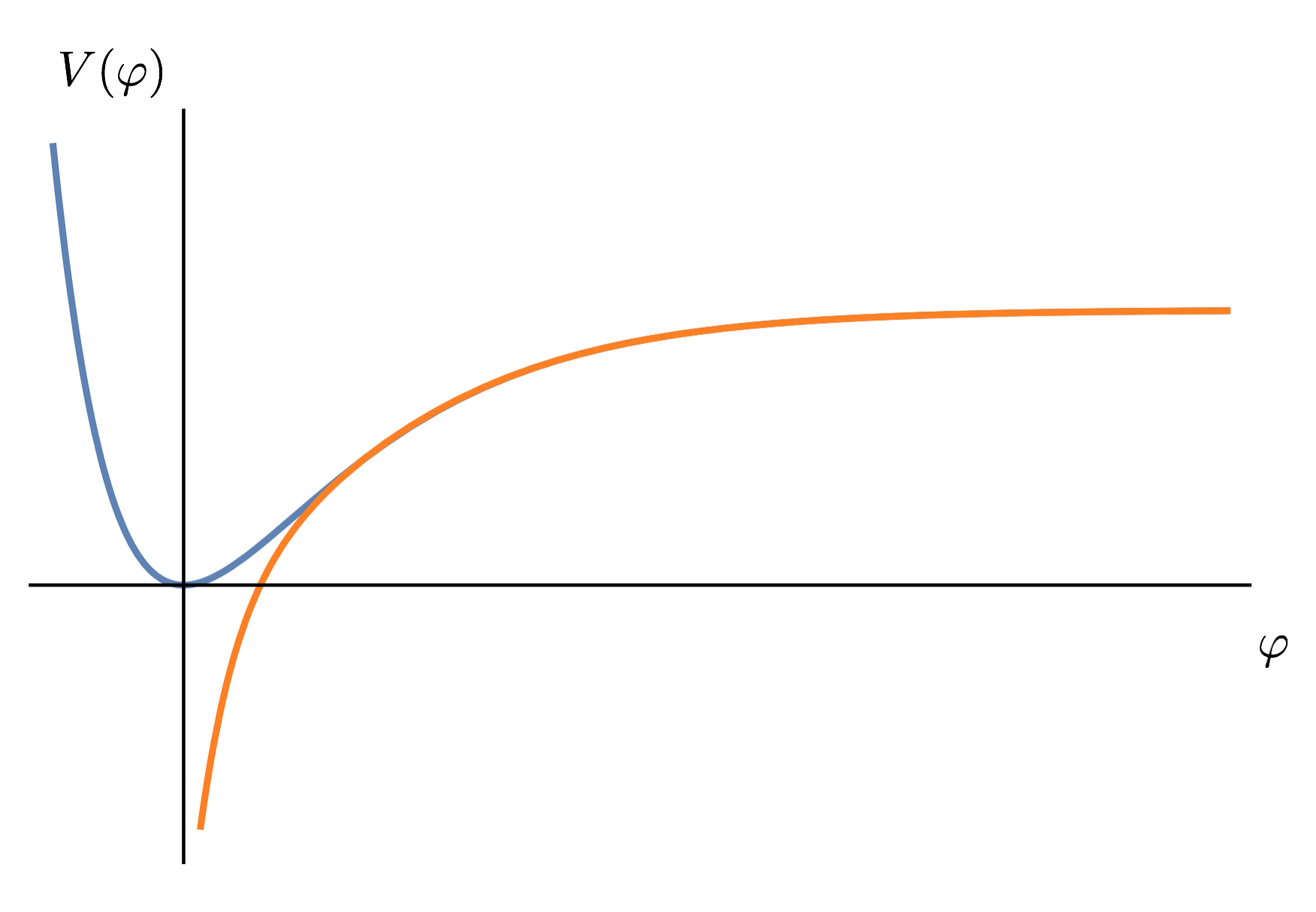} 
\caption{Schematic depiction of the Starobinsky potential potential (blue line) and the effective inflaton potential after integrating out $S$ (orange line), for $W_0 > M$. The theory does not have Minkowski vacua when the scale of supersymmetry breaking is this large, although a manipulation of the relation between $f$ and $W_0$ may produce de Sitter vacua with a large cosmological constant.}
\end{figure}
Since $M$ is fixed by observations this translates to a rigorous upper bound for the allowed scale of supersymmetry breaking,
\begin{equation}
m_{3/2} \lesssim 10^{12} \,{\rm GeV}\,.
\end{equation}
This is even lower than the upper bound found in \cite{Buchmuller:2014pla} for chaotic inflation. This is owed to the slightly lower Hubble scale in Starobinsky's inflation model. As illustrated in more detail in \cite{Dudas:2015lga}, in general we expect the dynamics of the stabilizer field to become important once the scale of supersymmetry breaking is comparable to the dynamical scale of inflation. In many examples, like this one, unfortunately, the backreaction becomes destructive beyond this point.

\section{Conclusion and outlook}

We have attempted to review a series of recent works on the backreaction of heavy moduli and stabilizer fields in supergravity inflation. While these backreactions are well under control and decouple for large mass hierarchies in the supersymmetric case, as investigated in \cite{Buchmuller:2014vda}, things are more complicated when supersymmetry is spontaneously broken by the heavy fields. For stabilized K\"ahler moduli which break supersymmetry, like in the LVS setup or uplifted KKLT, backreactions introduce soft terms which can drive inflation, but also other terms which may destabilize the inflationary trajectory at large field values. Therefore, stability of the system usually requires a very high scale of supersymmetry breaking in the vacuum. In models with a stabilizer field and broken supersymmetry, on the other hand, the requirement seems to be the opposite. Generically, supersymmetry breaking in the vacuum introduces a coupling between the inflaton and the stabilizer field which destabilizes the latter. The corresponding backreaction makes many otherwise successful inflation models unviable if the gravitino mass is comparable to the Hubble scale or larger.

We have illustrated our findings in two representative examples. For the backreaction of a K\"ahler modulus we chose chaotic inflation with a quadratic potential. In that case integrating out $T$ leads to a flattening of the potential, i.e., to an effective potential of the form $V = \frac12 m^2 \varphi^2 - \lambda \varphi^4$. There are viable regions of parameter space where these models predict CMB observables in good agreement with recent observations. This viability implies that $m_{3/2}$ must be close to the scale of Grand Unification. In this example the backreaction of $T$ helps to realize a successful model. This is not always the case. In \cite{Dudas:2015lga} there are different examples where the same backreaction prohibits inflation. As an example of the backreaction of stabilizer fields we chose a plateau model resembling the Starobinsky potential. There we observe that corrections induced by the dynamics of $S$ modify the vacuum structure in a way that the gravitino mass must be orders of magnitude below the Hubble scale for the model to be viable. Similar results have been found for other plateau-like models with a stabilizer field in \cite{Dudas:2015lga} and for chaotic inflation with a stabilizer field in \cite{Buchmuller:2014pla}.

Thus, our two examples are in this way distinct: One requires high-scale supersymmetry, the other low-scale supersymmetry. A combination of the two setups seems impossible to realize. This has implications for UV embeddings of inflation models with stabilizer fields, such as the ones considered in \cite{Dudas:2014pva,Escobar:2015fda}, once moduli stabilization is treated consistently. In many UV embeddings, like those involving supersymmetry breaking by moduli or fluxes, the scale of supersymmetry breaking is too high, thereby causing the stabilizer field to prohibit inflation.

\subsection*{Acknowledgements}

The work of CW has been supported by the ERC Advanced Grant SPLE under contract ERC-2012-ADG-20120216-320421, by the grant FPA2012-32828 from the MINECO, 
and by the grant SEV-2012-0249 of the ``Centro de Excelencia Severo Ochoa" Programme.

\end{document}